\documentclass[10.5pt,compsoc]{tst}
\usepackage{graphicx}
\usepackage{footmisc}
\usepackage{subfigure}
\usepackage{url}
\usepackage{multirow}
\usepackage[noadjust]{cite}
\usepackage{amsmath,amsthm}
\usepackage{amssymb,amsfonts}
\usepackage{booktabs}
\usepackage{color}
\usepackage{ccaption}
\usepackage{booktabs}
\usepackage{float}
\usepackage{fancyhdr}
\usepackage{caption}
\usepackage{xcolor,stfloats}
\usepackage{comment}
\graphicspath{{figures/}}
\usepackage{cuted}  
\usepackage{captionhack}
\usepackage{epstopdf}
\usepackage{subcaption}
\usepackage{cite}
\usepackage[ruled,linesnumbered]{algorithm2e}
\usepackage{booktabs}

\headevenname{\zihao{-5}{\textbf{\emph{Tsinghua Science and Technology, February}} 2018, 23(1): 000--000}}%
\headoddname{{\sf}\quad {\textbf{\emph{Game-theoretic Distributed Learning Approach for Heterogeneous-cost Task Allocation with Budget Constraints}}}}%

\newtheoremstyle{mystyle}{0pt}{0pt}{\normalfont}{1em}{\bf}{}{1em}{}
\theoremstyle{mystyle}
\renewcommand\figurename{Fig.~}

\newcommand{\nop}[1]{}

\makeatletter
\def\@cite#1#2{\textsuperscript{[{#1\if@tempswa, #2\fi}]}}
\renewcommand{\@biblabel}[1]{[#1]\hfill}
\makeatother

\begin{document}

\thispagestyle{empty}

\clearpage

\hyphenpenalty=50000

\makeatletter
\newcommand\mysmall{\@setfontsize\mysmall{7}{9.5}}

\newenvironment{tablehere}
  {\def\@captype{table}}
  {}
\newenvironment{figurehere}
  {\def\@captype{figure}}
  {}

\thispagestyle{plain}%
\thispagestyle{empty}%

\let\temp\footnote
\renewcommand \footnote[1]{\temp{\zihao{-5}#1}}
{}
\vspace*{-40pt}
\noindent{\zihao{5-}\textbf{\scalebox{0.885}[1.0]{\makebox[5.9cm][s]
{TSINGHUA\, SCIENCE\, AND\, TECHNOLOGY}}}}

\vskip .2mm
{\zihao{5-}
\textbf{
\hspace{-5mm}
\scalebox{1}[1.0]{\makebox[5.6cm][s]{%
I\hfill S\hfill S\hfill N\hfill{\color{white}%
l\hfill l\hfill}1\hfill0\hfill0\hfill7\hfill-\hfill0\hfill2\hfill1\hfill4
\hfill \color{white}{\quad 0\hfill ?\hfill /\hfill ?\hfill ?\quad p\hfill p\hfill  ?\hfill ?\hfill ?\hfill --\hfill ?\hfill ?\hfill ?}\hfill}}}}


\vskip .2mm\noindent
{\zihao{5-}\textbf{\scalebox{1}[1.0]{\makebox[5.6cm][s]{%
\color{white}{V\hfill o\hfill l\hfill u\hfill m\hfill%
e\hspace{0.356em}1,\hspace{0.356em}N\hfill u\hfill%
m\hfill b\hfill e\hfill r\hspace{0.356em}1,\hspace{0.356em}%
S\hfill e\hfill p\hfill t\hfill e\hfill%
m\hfill b\hfill e\hfil lr\hspace{0.356em}2\hfill0\hfill1\hfill8}}}}}\\

\begin{strip}
{\center
{\zihao{3}\textbf{
Game-theoretic Distributed Learning Approach for Heterogeneous-cost Task Allocation with Budget Constraints
\vskip 1.2mm\noindent
 }}
\vskip 9mm}

{\center {\sf \zihao{5}
Weiyi Yang, Xiaolu Liu, Lei He$^*$, Yonghao Du and Yingwu Chen
}
\vskip 5mm}

\centering{
\begin{tabular}{p{160mm}}

{\zihao{-5}
\linespread{1.6667} %
\noindent
\bf{Abstract:} {\sf
This paper investigates heterogeneous-cost task allocation with budget constraints (HCTAB), wherein heterogeneity is manifested through the varying capabilities and costs associated with different agents for task execution. Different from the centralized optimization-based method, the HCTAB problem is solved using a fully distributed framework, and a coalition formation game is introduced to provide a theoretical guarantee for this distributed framework. To solve the coalition formation game, a convergence-guaranteed log-linear learning algorithm based on heterogeneous cost is proposed. This algorithm incorporates two improvement strategies, namely, a cooperative exchange strategy and a heterogeneous-cost log-linear learning strategy. These strategies are specifically designed to be compatible with the heterogeneous cost and budget constraints characteristic of the HCTAB problem. Through ablation experiments, we demonstrate the effectiveness of these two improvements. Finally, numerical results show that the proposed algorithm outperforms existing task allocation algorithms and learning algorithms in terms of solving the HCTAB problem.}
}

\end{tabular}
}
\vskip 6mm

\vskip -3mm
\zihao{6}\end{strip}

\thispagestyle{plain}%
\thispagestyle{empty}%
\makeatother
\pagestyle{tstheadings}

\begin{figure}[b]
\vskip -6mm
\begin{tabular}{p{44mm}}
\toprule\\
\end{tabular}
\vskip -4.5mm
\noindent
\setlength{\tabcolsep}{1pt}
\begin{tabular}{p{1.5mm}p{79.5mm}}

$\bullet$&  Weiyi Yang, Xiaolu Liu, Lei He, Yonghao Du and Yingwu Chen are with the  College of systems engineering, National University of Defense Technology, Changsha 410000, China. E-mail:yangweiyi15@nudt.edu.cn, lxl\_sunny\_nudt@live.cn, helei@nudt.edu.cn, duyonghao15@163.com, ywchen@nudt.edu.cn. \\

$\sf{*}$&
To whom correspondence should be addressed. \\
          &          Manuscript received: 2024-4-5;
          accepted: 2024-6-13

\end{tabular}
\end{figure}\zihao{5}

\vspace{3.5mm}
\section{Introduction}
\label{s:introduction}
\noindent
Collaboration among multiple agents enables the execution of more complex tasks than can be achieved by a single agent, thereby enhancing the overall efficiency of the multi-agent system\cite{1}. In real life, numerous complex tasks require coordinated efforts from multiple agents, such as multi-resource disaster emergency rescue\cite{2}, coordination amongst multiple unmanned aerial vehicles (UAVs)\cite{3}, cooperative radar jamming\cite{4}, and collaborative imaging by remote sensing satellites\cite{5}. In these scenarios, heterogeneous agents incur varying costs when performing different tasks, and are typically subject to a predefined budget constraint, such as storage limits for imaging satellites or battery constraints for UAVs. Consequently, this paper addresses heterogeneous-cost task allocation with budget constraints (HCTAB).

 The complexity of the HCTAB problem mainly arises from the heterogeneity and large cardinality of the systems and agents under budget constraints. In the HCTAB problem, agents can only participate in at most one task, while each task may require the combined effort of multiple agents, i.e., the single-task robot and multi-robot task problem\cite{6}. Drawing inspiration from previous work\cite{7}, such problems can be viewed as a coalition formation problem, in which agents performing the same task automatically form a coalition. The coalition formation problem is NP-hard\cite{8}, and the computational complexity grows exponentially with the number of agents and tasks. Furthermore, the complexity escalates when considering agent heterogeneity under budget constraints. The heterogeneity of agents manifests in two aspects: differing competencies among agents and varying costs when executing different tasks. Specifically, agents are equipped with their own capabilities, and different tasks might require different capabilities. Additionally, the costs incurred by each agent vary when performing different tasks, adding a higher level of complexity to the efficient allocation of agents to tasks within a limited budget.

Considering the complexity and NP-hard characteristics of the HCTAB problem, an efficient distributed decision-making framework with superior robustness and scalability to a centralized approach is required\cite{9}. Unlike centralized allocation mechanisms, distributed allocation methods do not depend on a single main controller. Instead, agents collaborate and communicate with each other to collectively accomplish the global objective. However, achieving a high-quality solution without the aid of a centralized controller can be difficult. Therefore, the primary challenge in a distributed approach lies in designing individual decision-making algorithms that theoretically contribute to the global objective.

To solve this challenge, a novel distributed multi-agent task allocation framework for heterogeneous agents is designed based on the coalition formation game (CFG) and exact potential game (EPG). As a branch of game theory, CFGs are used to study the behavior of players when they engage in cooperative interactions\cite{10}, thus providing a practical analytical framework for solving the HCTAB problem in a distributed manner. Specifically, by designing the coalition value function, participant preferences, and coalition formation rules within the CFG, the global objective function of HCTAB can be effectively decomposed into local utility functions for each agent. This provides a theoretical foundation for individual-level decision-making within the multi-agent system, ensuring a stable coalition partition through the application of concepts and learning algorithms within the CFG.

Under budget constraints with heterogeneous-cost agents, we formulate the multi-agent task allocation problem into a CFG framework to achieve a convergence-guaranteed distributed task allocation. A log-linear learning algorithm based on heterogeneous cost (LLH) is then proposed to solve the game. To the best of our knowledge, this is the first attempt to address the HCTAB problem from a CFG-theoretic perspective and provide a distributed, convergence-guaranteed approach for its efficient solution via local coordination alone. The main contributions of this work are as follows:

\begin{itemize}
    \item Considering the complexity of centralized computation, the proposed LLH aims to solve the HCTAB problem in a fully distributed framework. A CFG is introduced to model the HCTAB and provide a theoretical guarantee for this distributed framework, where the coalition value function, participant preference order, and coalition formation rules are customized according to the HCTAB characteristics. Next, the convergence to a stable coalition structure is proved by converting the CFG into an EPG.
    \item The proposed LLH is designed to solve the constructed CFG. Two strategies are designed to incorporate heterogeneous costs and budget constraints: a cooperative exchange (CE) strategy and a heterogeneous-cost log-linear learning (HLL) strategy. Our LLH formulation  is computationally efficient, distributed, and highly scalable, making it a suitable candidate for large-scale allocation problems.
    \item Numerical experiments are conducted with different numbers of agents, degrees of heterogeneity, and budget rates. The results of these experiments demonstrate the superiority of the LLH algorithm, in terms of both solution quality and efficiency, over state-of-the-art algorithms.
\end{itemize}
The remainder of this paper is organized as follows. Section 2 discusses related work on task allocation problems. Section 3 defines the task allocation problem with budget constraints and introduces the game formulation for such problems. Section 4 presents the proposed LLH algorithm, before Section 5 demonstrates the efficiency of the CE and HLL algorithmic strategies in numerical experiments, and presents comparisons with recent state-of-the-art algorithms. Finally, Section 6 summarizes the conclusions from this study and outlines directions for future work.

\section{Related work}
\label{s:Problem description}
\noindent
From the perspective of multi-agent collaboration, existing methods can be categorized into two main types: centralized task allocation and distributed task allocation. Centralized task allocation requires a central controller to monitor the real-time status of all tasks and agents, and to allocate all tasks to all agents accordingly. Under this approach, allocation algorithms can be classified into exact solution algorithms and metaheuristic methods. Exact solution algorithms, such as the Hungarian algorithm\cite{11}, branch-and-bound algorithm\cite{12}, and dynamic programming\cite{13}, aim to produce high-quality solutions in small-scale task allocation or coalition formation problems by employing effective branching, pruning, or search strategies. For example, Michalak et al.\cite{14} proposed a hybrid exact algorithm that combines an optimal version of dynamic programming and a tree-search algorithm to identify the optimal coalition partition. However, as the problem scale increases and the constraints become more complex, exact solution algorithms struggle to find optimal solutions within an acceptable runtime\cite{9}.

Other centralized task allocation approaches include optimization-based methods, represented by local search algorithms\cite{15}, swarm intelligence algorithms\cite{16}, and bio-inspired evolutionary algorithms\cite{17,18}, which have recently emerged as a promising approach for multi-agent allocation problems. For example, Amorim et al.\cite{19} solved dynamic allocation problems incorporating multiple agents using a swarm intelligence strategy, while Yan et al.\cite{20} presented a new hyper-heuristic algorithm for multi-agent task allocation. Although optimization-based methods offer excellent exploration performance and are widely used, it is hard to design appropriate local decision rules\cite{21}. Thus, their superiority disappears when applied to HCTAB problems in distributed systems.

Market-based approaches (MBAs)\cite{22}, developed from the contract net protocol\cite{23}, are another competitive means of dealing with the distributed scenario of this problem. In MBAs, multiple bids are made by agents and compared to determine the best offer, with the final deals going to the highest bidders. For example, Lee et al.\cite{24} proposed a resource-oriented, decentralized auction algorithm for multi-robot task allocation, and later compared the proposed auction-based algorithm with other MBAs\cite{7}. Another MBA for task allocation problems is the consensus-based approach, which associates an auction phase or a bundle construction phase with a consensus phase\cite{25} to give the consensus-based auction algorithm (CBAA) or the consensus-based bundle algorithm (CBBA), respectively. CBAA allocates at most one task per agent, whereas CBBA allocates multiple tasks per agent. Thus, CBBA is an extension of CBAA that incorporates multi-assignment. Ye et al.\cite{26} developed an extended CBBA that reaches a global consensus and obtains conflict-free task assignment results in a heterogeneous agent system. However, without the aid of a centralized controller, it is difficult for MBAs to theoretically guarantee that the local decisions of the individual benefit the whole system, which limits the solution quality.

Game-theoretic formulations provide an alternative angle for the modeling of task allocation problems. By regarding the agents allocated to the same task as a coalition, we can transform the original HCTAB problem into a CFG to facilitate problem-solving. CFGs are used in game theory to study the behavior of players when they cooperate among themselves, and have been widely applied in ad-hoc network clustering\cite{27}, UAV network clustering\cite{28}, and cooperation among roadside units\cite{29}. Similarly, CFGs have been extensively used in solving multi-agent task allocation problems. For example, Chen et al.\cite{30} formulated the joint task and spectrum allocation problem as a CFG, using the concept of Nash stability to determine an equilibrium coalition structure. However, this formulation did not consider the budget feature in HCTAB. Li et al.\cite{31} incorporated system budget constraints into the task allocation problem, modeling it as a CFG, and proposed a greedy-based iterative algorithm to search for a stable coalition. In subsequent work\cite{7}, they further introduced a cost-efficient factor to guide the formation of the coalition. However, they assumed that each agent has the same cost for performing any non-dummy tasks, which may not be realistic in real-world situations.

\section{Problem formulation and game modeling}
\noindent

\subsection{Problem formulation}
\noindent
The focus of this paper revolves around the HCTAB problem, taking into account task execution by a specific set of agents and tasks that embody multiple complexities. Suppose that there exist a set of $n$ agents $N=\left\{ 1,2,\ldots ,i\ldots ,n \right\}$  and a set of $m$ tasks $T=\left\{ 1,2,\ldots ,j\ldots ,m \right\}$. An agent cannot perform multiple tasks at the same time, and each task can be executed by several agents. Some tasks might not be feasible for certain agents. ${{T}_{i}}$ represents the feasible task set for agent $i$ under the spatial and temporal constraints.
.
Heterogeneity remains at the core of the HCTAB problem, with both the agents and the tasks considered to have heterogeneous capacities. Let $W=\left\{ 1,\ldots,k\ldots,l \right\}$ be the capacity set of the allocation system. Each agent $i$ possesses a subset of capacities, ${{w}_{i}} \in W$. These capacities have different levels of competencies, denoted as ${{h}_{ik}}$. The tasks also require a diverse set of capacities, represented by ${{w}_{j}}\in W$ for each task $j$.

Let ${{S}_{j}}$ denote the set of agents allocated to task $j$. In terms of the rewards linked with task execution, this study contemplates a model consistent with that in\cite{7}, formulating the reward ${{R}_{j}}$ received for task $j$ as an accumulative sum of maximum competencies across its allocated agents:
\begin{equation}
    {{R}_{j}}=\sum\limits_{k\in {{w}_{j}}}{\underset{i\in {{\{S\}}_{j}}}{\mathop{\max }}\,{{h}_{ik}}}
\end{equation}

The aspect of heterogeneity in HCTAB is further emphasized through the varying costs incurred by different agents for undertaking diverse tasks. This heterogeneity is represented through a unique cost ${{c}_{ij}}$ when agent $i$ is assigned task $j$. The total cost across all agents and tasks is limited by a predefined budget constraint $B$.

The objective of the task allocation problem is to find the optimal assignment that maximizes the sum of task rewards. The single-task reward can be formulated as Equation (1). However, the work capacity of all agents is finite, and it is necessary to make full use of a limited budget to ensure that the task-reward system operates optimally. Therefore, the HCTAB problem can be formulated as follows.
\begin{equation}
    \max \sum\limits_{j\in T}{{{R}_{j}}}=\sum\limits_{k\in {{w}_{j}}}{\underset{i\in N}{\mathop{\max }}\,{{h}_{ik}}{{x}_{ij}}}
\end{equation}
subject to:
\begin{equation}
    \sum\limits_{j\in T}{{{x}_{ij}}}\le 1\quad \forall i\in N
\end{equation}
\begin{equation}
    \sum\limits_{i\in N}{\sum\limits_{j\in T}{{{c}_{ij}}{{x}_{ij}}}}\le B
\end{equation}

where ${{x}_{ij}}$ is a binary decision variable that indicates whether agent $i$ is assigned to task $j$, and $\sum\limits_{i\in N}{\sum\limits_{j\in T}{{{c}_{ij}}{{x}_{ij}}}}$ is the total cost of the allocation profile, which cannot exceed the budget $B$.

\subsection{Coalition formation game }
\noindent
Emerging from cooperative game theory, CFGs have recently attracted considerable attention, offering significant theoretical backing to cooperative task allocation while distributing the ensuing rewards among the participants. In this section, we model the HCTAB problem as a CFG with transferable utility (TU)\cite{32}. Each agent in HCTAB is interpreted as a participant in a CFG, where agents performing the same task form coalitions and the remaining agents without executable tasks form a dummy coalition. Let $\Pi ~=\text{ }\left\{ {{C}_{1}},...,\text{ }{{C}_{m}} \right\}$  represent a coalition structure, where each ${{C}_{j}}\in \Pi $ indicates the coalition for task $j$. An agent can join at most one coalition according to the HCTAB problem description. The values in TU games allow members of a coalition to assess their contributions to this coalition using an appropriate fairness rule. To derive our distributed task allocation approach, we recall some definitions from coalition game theory, starting with the notions of the coalition structure preference relation and the CFG itself.

\textbf{Definition 1} (\textit{Coalition structure}): A coalition structure (or coalition partition) is defined as the set $\Pi:=\left\{ {{C}_{1}},...,{{C}_{m+1}} \right\}$, where ${{C}_{j}}\subseteq N$ are disjoint coalitions satisfying $\bigcup\nolimits_{j=1}^{m+1}{{{C}_{j}}=N}$.

\textbf{Definition 2} (\textit{Coalition formation game}): A CFG is given by a pair ($N$, $V$), where $N$ is the set of game participants and $V$ is a function over the real line. The value that coalition ${{C}_{j}}\in \Pi $ can receive is quantified as $v({{C}_{j}})$. 

The value $v({{C}_{j}})$ thus serves as the total utility engendered by the interaction of agents within coalition ${{C}_{j}}$. This illustrates the fundamental premise of a CFG, where participants unite in diverse combinations to maximize their collective utility. Three different quantities are introduced to each coalition ${{C}_{j}}$: the revenue function $r({{C}_{j}})$, the cost function $c({{C}_{j}})$, and the coalition value $v({{C}_{j}})$.

$\bullet$ The revenue function $r({{C}_{j}})$ quantifies the worth of the coalition. In this study, we define the revenue function for coalition $j$ as the reward for task $j$: 
    \begin{equation}
        r({{C}_{j}})={{R}_{j}}=\sum\limits_{k\in {{W}_{j}}}{\underset{i}{\mathop{\max }}\,{{h}_{ik}}}
    \end{equation}

$\bullet$ The cost function $c({{C}_{j}})$ quantifies the cost of cooperation. The cost function is used to account for the constraint:
    \begin{equation}
        c({{C}_{j}})=\left\{ \begin{matrix}
        0\quad \quad if\ \sum\limits_{j\in T}{{{x}_{ij}}}\le 1,\ \sum\limits_{i\in N}{\sum\limits_{j\in T}       {{{c}_{ij}}{{x}_{ij}}}}\le B  \\
        +\infty \quad \quad \quad \quad \quad \quad \quad \quad otherwise  \\
        \end{matrix} \right.
    \end{equation}

$\bullet$ The value $v({{C}_{j}})$ is defined as the difference between $r({{C}_{j}})$ and $c({{C}_{j}})$ as follows:
    \begin{equation}
        v({{C}_{j}})=r({{C}_{j}})-c({{C}_{j}})
    \end{equation}

The value function represents the utility received by a coalition. By applying the TU property, the utility of the coalition can be arbitrarily divided between the coalition members. According to the wonderful-life utility\cite{33}, which decomposes the coalition-level utility to each agent, the agent utility ${{u}_{i}}({{C}_{j}})$ is defined as the marginal contribution of agent $i$ to its coalition:
\begin{equation}
    {{u}_{i}}({{C}_{j}})=v\left( {{C}_{j}} \right)-v\left( {{C}_{j}}/\{i\} \right)
\end{equation}

\subsection{Coalition formation game elements}
\noindent
The inherent complexity of CFGs necessitates efficient strategies for their optimal solution. Centralized methods, while exploring cost--benefit tradeoff cooperation, are limited by their NP-complete nature, leading to exponentially increasing iteration times. In the previous subsection, we decomposed the coalition value function into the utility functions of each agent. This helps agents make decisions in a distributed manner based on the current state, effectively reducing the problem dimension and complexity. Before designing distributed coalition formation algorithms, three elements need to be determined and analyzed according to the characteristics of the HCTAB problem: \textit{preference orders}, \textit{coalition formation rules}, and \textit{coalition stability}.

\subsubsection{Preference orders}
\noindent
All agents can be allocated to any feasible task that satisfies the budget constraints, resulting in the formation of a coalition. However, task rewards are typically variable depending on the specific coalition (or task) an agent elects to join. Consequently, agents demonstrate different preferences over different coalitions. 

\textbf{Definition 3} (\textit{Preference order}): For any participant $i\in N$, a preference order ${{\succ }_{j}}$ is defined as a complete, reflexive, and transitive binary relation over the set of all coalitions that participant $i$ can possibly join.

The ordinary preference order includes the selfish preference and Pareto preference. Conversely, under the Pareto order, agents are obligated to ensure an enhanced individual utility for every agent in both the original and new coalitions. However, these rigorous constraints make it challenging for agents to leave the original coalition and join a new one, rendering it unsuitable for our HCTAB problem, which includes multiple complexities and a high degree of heterogeneity. To counter this issue, a coalition preference order is implemented. In the context of the selfish order, agents solely contemplate their own utility increment whilst disregarding the utilities of other agents.

\textbf{Definition 4} (\textit{Coalition preference order}): If any two potential coalitions can be joined by participant $i\in N$, i.e., ${{C}_{j}}\in \Pi $ and ${{C}_{{{j}'}}}\in \Pi $, then the preference relation is as follows:
\begin{equation}
    \begin{aligned}
    & {{C}_{j}}{{\succ }_{i}}{{C}_{{{j}'}}}\Leftrightarrow  \\ 
    & {{u}_{i}}\left( {{C}_{j}} \right)+v\left( {{C}_{j}}\cup \left\{ i \right\} \right)-v\left( {{C}_{j}} \right)> \\ 
    & {{u}_{i}}\left( {{C}_{{{j}'}}} \right)+v\left( {{C}_{{{j}'}}}\cup \left\{ i \right\} \right)-v\left( {{C}_{{{j}'}}} \right) \\ 
    \end{aligned}
\end{equation}

This implies that agent $i$ will choose to join the coalition that enhances the collective utility of itself, as well as the original and new coalitions. In contrast to Pareto ordering, the coalition preference order only considers the aggregate utility of members within its own coalition, rather than individually considering the utilities of all members in the coalition.

\subsubsection{Coalition formation rules}
\noindent
The rules governing coalition formation and dissolution among nodes are now defined. The coalition selection rule was initially proposed by Ruan et al.\cite{34}, offering simplicity, comprehensibility, and efficient guidance for coalition formation. 

\textbf{Definition 5} (\textit{Coalition selection rule}\cite{34}): Given two coalitions ${{C}_{j}}\in \Pi $ and ${{C}_{{{j}'}}}\in \Pi $, an arbitrary agent $i\in N$ prefers to join coalition ${{C}_{j}}$ if $i$ prefers to be part of ${{C}_{j}}$ over being part of ${{C}_{{{j}'}}}$ based on the coalition preference order, i.e.,
\begin{equation}
    i\to {{C}_{j}}\Leftrightarrow {{C}_{j}}{{\succ }_{i}}{{C}_{j'}}\quad \forall {{C}_{j}},{{C}_{j'}}\in \Pi
\end{equation}

The coalition selection rule provides a mechanism through which agent $i$ can leave the original coalition and select a new coalition to join if and only if the new coalition is preferred over the current coalition, which is defined in Equation (9). Based on the coalition selection rule, agent $i$ will join the coalition that increases the total utility of the allocation system.

\subsubsection{Analysis of the stable coalition partition}
\noindent
We now analyze the stability of the partitions in our CFG. The concept of stability in the game is defined as follows.

\textbf{Definition 6} (\textit{Stable coalition partition}): A coalition partition $\Pi ~=\text{ }\left\{ {{C}_{1}},...,\text{ }{{C}_{m}} \right\}$ is stable if there is no participant $i\in N$ with a preference for other coalitions. In other words, there is no participant who can improve its utility by changing its coalition (task selection) unilaterally. That is,
\begin{equation}
    {{u}_{i}}({{a}_{i}},{{a}_{-i}})\ge {{u}_{i}}({{{a}'}_{i}},{{a}_{-i}})\quad \forall i\in N,\forall {{a}_{i}},{{{a}'}_{i}}\in {{A}_{i}}
\end{equation}

where ${{a}_{i}}$ exemplifies the action of participant $i$, denoting the coalition with which agent $i$ selects to be aligned. If agent $i$ selects coalition ${{C}_{j}}$ (i.e., allocated to task $j$), ${{a}_{i}}$ is equal to ${{C}_{j}}$. The collection of ${{a}_{i}}$ is denoted as ${{A}_{i}}$ (${{A}_{i}}=T\cup \{m+1\}$), where coalition ($m$+1) contains all unsigned agents and the value of coalition ($m$+1) is equal to zero according to Equation (5). Here, ${{a}_{-i}}$ represents the coalition selections of agents other than agent $i$.

In contrast to conventional preference orders, the coalition expected order considers the interests of other participants. To examine stability under the coalition expected order, we introduce the concept of an EPG to analyze partition stability under the coalition preference order.

\textbf{Definition 7} (\textit{Exact potential game}): A game is an EPG when there exists a potential function $\varphi:{{A}_{1}}\times ...\times {{A}_{n}}\to R$  satisfying
\begin{equation}
    \begin{aligned}
   & {{u}_{i}}({{{a}'}_{i}},{{a}_{-i}})-{{u}_{i}}({{a}_{i}},{{a}_{-i}})=\varphi ({{{a}'}_{i}},{{a}_{-i}})-\varphi ({{a}_{i}},{{a}_{-i}})\quad \\
   & \forall {{\tilde{a}}_{i}},{{a}_{i}}\in {{A}_{i}}
    \end{aligned}
\end{equation}

\textbf{Theorem 1}: Under the coalition preference order and utility function outlined in Equation (8), the CFG can converge to a stable coalition partition. The optimal solution to the HCTAB problem constitutes a stable coalition partition.

\textbf{Proof}: As introduced in Ref. [33], we denote the potential function as
\begin{equation}
    \varphi ({{a}_{i}},{{a}_{-i}})=\sum\limits_{j\in T\cup \{m+1\}}{v({{C}_{j}})}
\end{equation}
which represents the sum of all coalition values and is equivalent to the objective function of the HCTAB problem. Suppose that agent $i$ takes action ${{a}_{i}}$ in selecting coalition ${{C}_{j}}$ and takes action ${{a}_{{{i}'}}}$ in selecting coalition ${{C}_{{{j}'}}}$. The original coalition to which agent $i$ belongs is denoted as ${{C}_{{\hat{j}}}}$. The change in the potential function resulting from this individual action adjustment is calculated as follows:
\begin{equation}
    \begin{aligned}
  & \varphi ({{a}_{i}},{{a}_{-i}})-\varphi ({{{{a}'}}_{i}},{{a}_{-i}}) \\ 
 & =v\left( {{C}_{j}}\cup \left\{ i \right\} \right)+v\left( {{C}_{{\hat{j}}}}/\left\{ i \right\} \right)+\sum\limits_{\bar{j}\in T/\{j,\hat{j}\}}{v\left( {{C}_{{\bar{j}}}} \right)} \\ 
 & -\left( v\left( {{C}_{{{j}'}}}\cup \left\{ i \right\} \right)+v\left( {{C}_{{\hat{j}}}}/\left\{ i \right\} \right)+\sum\limits_{\bar{j}\in T/\{{j}',\hat{j}\}}{v\left( {{C}_{{\bar{j}}}} \right)} \right) \\ 
 & =v\left( {{C}_{j}}\cup \left\{ i \right\} \right)+v\left( {{C}_{{{j}'}}} \right)+\sum\limits_{\bar{j}\in T/\{j,\hat{j},{j}'\}}{v\left( {{C}_{{\bar{j}}}} \right)} \\ 
 & -\left( v\left( {{C}_{{{j}'}}}\cup \left\{ i \right\} \right)+v\left( {{C}_{j}} \right)+\sum\limits_{\bar{j}\in T/\{{j}',\hat{j},j\}}{v\left( {{C}_{{\bar{j}}}} \right)} \right) \\ 
\end{aligned}
\end{equation}

In accordance with the designed coalition selection rule, the action taken by agent $i$ will not alter the coalition value of any coalition except ${{C}_{j}}$, ${{C}_{{{j}'}}}$, and ${{C}_{{\hat{j}}}}$. Consequently, the change in the potential function can be reformulated as
\begin{equation}
    \begin{aligned}
  & \varphi ({{a}_{i}},{{a}_{-i}})-\varphi ({{{{a}'}}_{i}},{{a}_{-i}})= \\ 
 & v\left( {{C}_{j}}\cup \left\{ i \right\} \right)+v\left( {{C}_{{{j}'}}} \right)-\left( v\left( {{C}_{{{j}'}}}\cup \left\{ i \right\} \right)+v\left( {{C}_{j}} \right) \right) \\ 
\end{aligned}
\end{equation}

If agent $i$ changes coalition from ${{C}_{j}}$ to ${{C}_{{{j}'}}}$, given the utility function defined in Equation (8), the difference in utility can be expressed as
\begin{equation}
    \begin{aligned}
  & {{u}_{i}}({{a}_{i}},{{a}_{-i}})-{{u}_{i}}({{{{a}'}}_{i}},{{a}_{-i}})= \\ 
 & v\left( {{C}_{j}}\cup \left\{ i \right\} \right)-v\left( {{C}_{j}} \right)-\left( v\left( {{C}_{{{j}'}}}\cup \left\{ i \right\} \right)-v\left( {{C}_{{{j}'}}} \right) \right) \\ 
 & =\varphi ({{a}_{i}},{{a}_{-i}})-\varphi ({{{{a}'}}_{i}},{{a}_{-i}}) \\ 
\end{aligned}
\end{equation}

According to Definition 7, the CFG is an EPG, which must contain at least one pure strategy Nash equilibrium\cite{33}. For the Nash equilibrium point $({{\tilde{a}}_{i}},{{\tilde{a}}_{-i}})$, the inequality ${{u}_{i}}({{\tilde{a}}_{i}},{{\tilde{a}}_{-i}})\ge {{u}_{i}}({{a}_{i}},{{\tilde{a}}_{-i}})$ holds for any agent $i$ and any action ${{a}_{i}}$ from ${{A}_{i}}$. Therefore, with the designed coalition preference order and coalition selection rule, the CFG converges to a stable coalition partition based on Definition 6. 

Moreover, from Equations (2) and (13), maximizing the potential function $\varphi$ is equivalent to maximizing the objective function of the HCTAB problem. The optimal solution to the HCTAB problem implies that the objective value cannot be improved by any actions, signifying the existence of a stable coalition partition in this game. Thus, Theorem 1 has been proved.

\section{Algorithm design}
The preceding section formulated the HCTAB problem as a CFG and EPG to decompose the global objective function into local utility functions for individual agents. Consequently, agents can update their actions based on their current state and updated action, adhering to their respective local utility functions. In the field of the potential game action learning approach, numerous learning algorithms such as fictitious play\cite{35}, best response\cite{36}, better reply process\cite{37}, and others\cite{9} have demonstrated strong convergence properties. Among them, log-linear learning\cite{38} guarantees that the action profile converges to a potential maximizer in potential games. Motivated by this, we now describe the proposed LLH. Two strategies are designed to incorporate heterogeneous costs and budget constraints: the CE strategy and the HLL strategy.

\begin{figure}[!htb]
    \centering
    
    \subfigure[Situation with only coalition selection]{
        \includegraphics[scale=0.38]{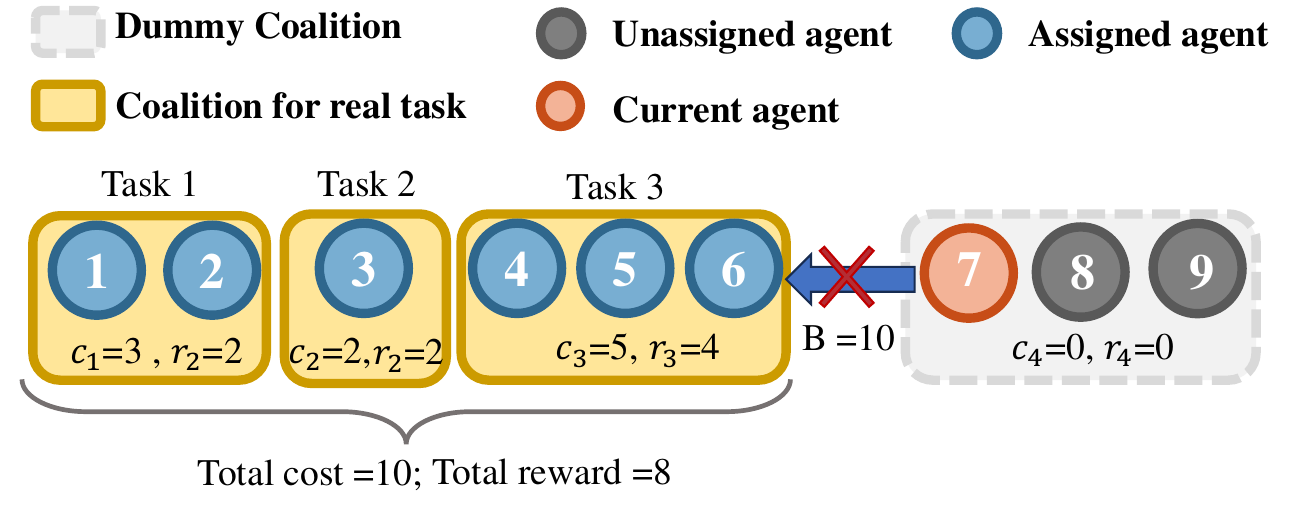}
    \label{F1A}
    }
    \quad
    \subfigure[Situation with the addition of the CE strategy]{
        \includegraphics[scale=0.38]{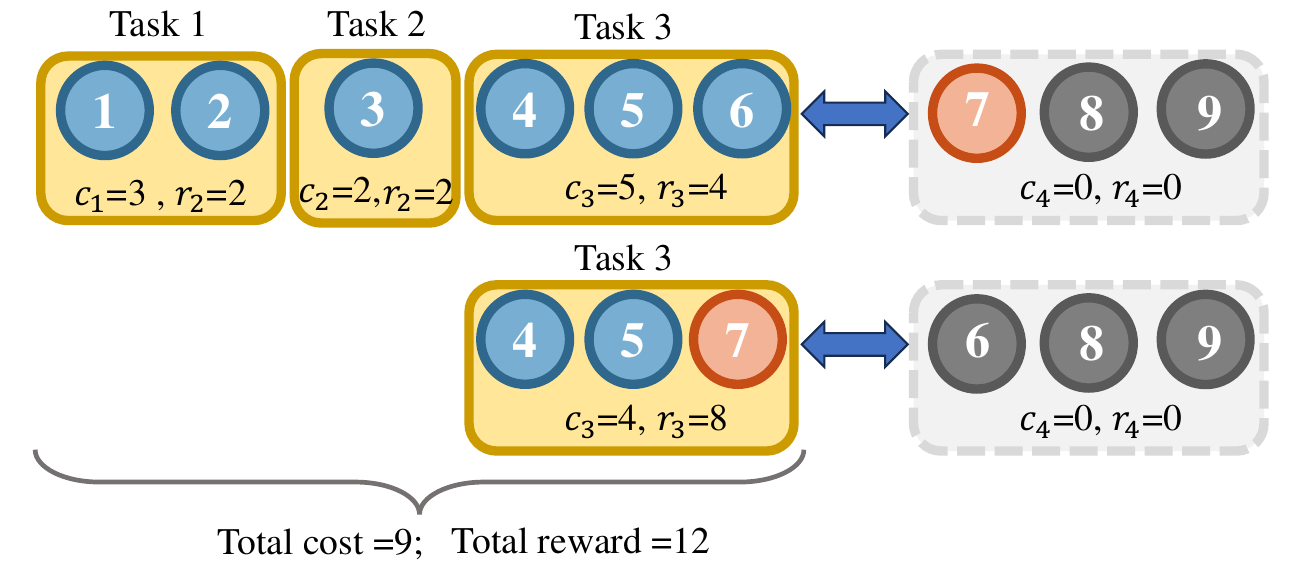}
    \label{F1B}
    }
\caption{ Example of the CE strategy.}
\label{f1}
\end{figure}

The CE strategy serves as a complementary approach to the coalition selection rule, aiming to expand the action space available for agents while considering the budget constraint in HCTAB. Under the designated coalition order and coalition selection rule, once an agent joins a coalition from an empty coalition ($m$+1) for a specific task $j$, that agent can only choose to transit from current task $j$ to another task $j'$, and is not allowed to return to the empty coalition ($m$+1). Similarly, when the total cost in the system reaches the budget value, unassigned agents are unable to directly join coalitions through the coalition selection rule. As depicted in Fig. \ref{f1}(a), there exist three coalitions comprising six agents with a total cost reaching the budget value of 10. Therefore, if agent 7 joins a coalition, it will further escalate the costs and violate the budget constraint. Figure \ref{f1}(b) illustrates how the CE strategy addresses this situation: unassigned agent 7 exchanges positions with assigned agent 6 without contravening any constraints and achieves higher overall benefits.

Under the coalition selection rule, it becomes challenging for assigned agents to revert to a dummy coalition. In a budget-constrained problem like HCTAB, the outcome of the algorithm will significantly rely on the sequence in which agents are assigned to tasks. However, in distributed algorithms, this assignment sequence is uncertain, which could directly impact the algorithm's efficiency. Therefore, the CE strategy enables unassigned agents to be exchanged for assigned agents that have lower cost efficiency.

\textbf{Definition 8} (\textit{Cooperative exchange strategy}): Given two coalitions ${{C}_{j}}\in \Pi $  and  ${{C}_{{{j}'}}}\in \Pi $, arbitrary participants $i,{i}'\in N$ prefer to exchange their coalitions based on the coalition altruistic preference order, i.e.,

\begin{equation}
    \begin{aligned}
  & i\to {{C}_{j}},{i}'\to {{C}_{{{j}'}}}\Leftrightarrow {{u}_{i}}\left( {{C}_{j}} \right)+{{u}_{i'}}\left( {{C}_{j'}} \right)\\ 
 & +v\left( {{C}_{j}}\cup \left\{ i \right\}/\{i'\} \right)+v\left( {{C}_{j'}}\cup \left\{ i' \right\}/\{i\} \right) \\ 
 & >{{u}_{i'}}\left( {{C}_{j}} \right)+{{u}_{i}}\left( {{C}_{j'}} \right)+v\left( {{C}_{j}} \right)+v\left( {{C}_{j'}} \right)\quad \\ 
 &  \forall i\in {{C}_{j'}},\ {i}'\in {{C}_{j}}
\end{aligned}
\end{equation}

The CE strategy offers two distinct advantages. First, it enables the flexible exchange of assigned agents with lower reward-cost efficiency by returning them to an unassigned state. This ensures that the algorithm is no longer strictly dependent on the assignment sequence of agents. Second, a single CE is equivalent to a combination of two coalition selections, effectively accelerating the convergence and enhancing the task allocation efficiency without impacting the conclusion of Theorem 1.

The HLL strategy is an action selection strategy considering heterogeneous costs. Ordinary log-linear learning chooses action ${{a}_{i}}$ with probability
\begin{equation}
    {{p}_{{{a}_{i}}}}=\frac{{{e}^{\beta {{u}_{i}}({{a}_{i}},{{a}_{-i}})}}}{\sum\limits_{{{{\bar{a}}}_{i}}\in {{A}_{i}}}{{{e}^{\beta {{u}_{i}}({{{\bar{a}}}_{i}},{{a}_{-i}})}}}}
\end{equation}

where the fixed parameter $\beta$ constrains the performance, because maximizing the potential function requires $\beta$ to approach infinity. Hence, we adopt a time-variant heterogeneous-cost parameter $\beta \left( t,\Delta c \right)$ to replace the fixed parameter $\beta$: 
\begin{equation}
    \beta \left( t,\Delta c \right)={{\beta }_{0}}\cdot \frac{\Delta c}{\Delta {{c}_{\max }}}+\frac{\ln \left( \lambda t+1 \right)}{c}
\end{equation}
where $\Delta c$ is the cost decrease caused by the current action at iteration $t$, $\Delta {{c}_{\max }}$ is the maximum possible value of $\Delta c$, and $\lambda $, ${{\beta }_{0}}$, $c$ are constants that satisfy ${{\beta }_{0}}\ge 0$, $\lambda \ge 1$, and $c\in {{N}^{*}}$. The convergence of log-linear learning with this variant-parameter has previously been proved\cite{9}.

The pseudocode for the proposed LLH strategy is given in Algorithm 1. Lines 2--15 represent the construction of the agent action space, where lines 2--6 indicate the actions constructed by the coalition selection rule and lines 7--15 represent the action space supplemented by the CE strategy. After completing the construction of the action space, agent $i$ selects action ${{a}_{i}}$ to be updated in the current iteration guided by the HLL strategy, as shown in line 17.

\begin{algorithm}[]\label{alg:1}
	\caption{Log-linear learning algorithm based on heterogeneous cost}
	\LinesNumbered
	\KwIn{current node $i$, current iteration $t$, current coalition partition $\Pi (t)$, feasible task set ${{T}_{i}}$, current coalition $\hat{j}$.}
	\KwOut{ action $a_{i}^{t}$ for agent $i$}
	Initialize the log-linear learning parameter $\beta$ and the possible action set ${{A}_{i}}=\varnothing $ \\
	\For{each feasible task $j\in {{T}_{i}}$}{
        \If{${{C}_{j}}{{\succ }_{i}}{{C}_{{\hat{j}}}}$}{
            ${{A}_{i}}={{A}_{i}}\cup \left\{ j \right\}$
        }
	}
    \If{${{A}_{i}}=\varnothing $}{
        \For{$j\in {{T}_{i}}$ and ${{C}_{j}}\ne \varnothing $}{
            \For{each agent ${{i}^{e}}\in {{C}_{j}}$}{
                \If{$i\to {{C}_{j}},{i}'\to {{C}_{{{j}'}}}$}{
                    ${{A}_{i}}={{A}_{i}}\cup \left\{ j \right\}$
                }
            }
        }
    }
    \If{${{A}_{i}}\ne \varnothing $}{
         Find the update action $a_{i}^{t}$ with probability in Eq. (19)
    }
	\Return{update action $a_{i}^{t}$}
\end{algorithm}

\begin{figure}[!htb]
    \centering
    \includegraphics[scale=0.48]{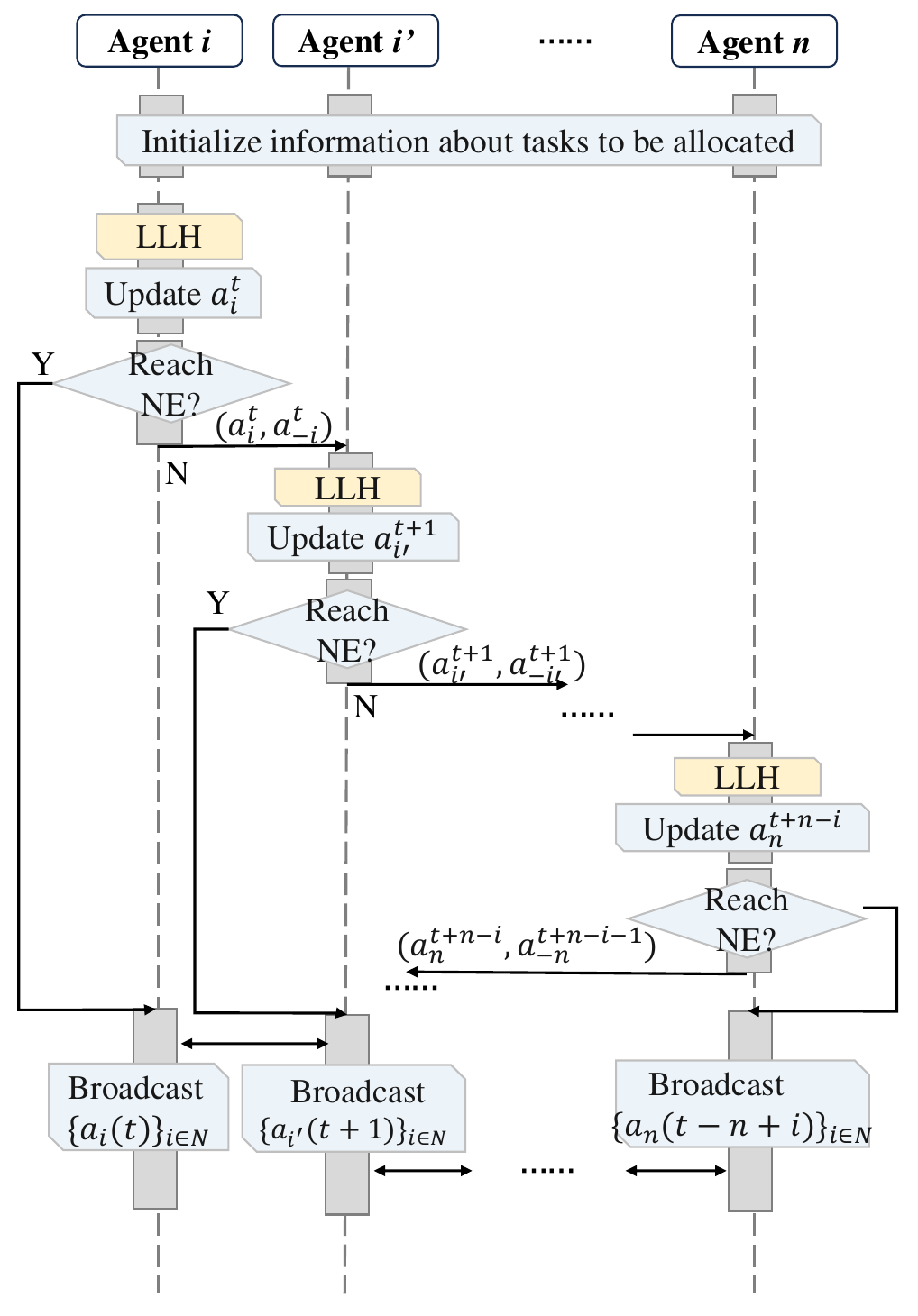}  
\caption{Diagram of the operational process of distributed methods in agent systems.}   
\label{f2}
\end{figure}

The operational process of distributed methods in agent systems is now introduced, with a smoke signal play\cite{39} employed to reduce the communication frequency among agents. Specifically, as depicted in Fig. \ref{f2}, all agents initialize their task information, including the required capability information and feasible agent set. The distributed task allocation process is then implemented in a relay manner, whereby agent $i$ receives allocation file $\left( a_{i}^{t}, a_{-i}^{t} \right)$\footnote{The coalition partition file and task allocation file are mutually equivalent. For convenience, the term `allocation file' refers to both the task allocation file and agent coalition partition file.} and the current system state (e.g., the total current cost) from another agent at iteration $t$. Based on the received information, each agent updates its action (i.e., task allocation file) by executing the LLH algorithm. Once the updated allocation file reaches the Nash equilibrium or the maximum number of iterations ${{T}_{\max }}$, the agent that receives the allocation scheme broadcasts the final allocation file globally.

\section{Experiments}
\noindent
We follow the experimental scenario settings from a previous study\cite{7}: the number of agents in the experiment is three times the number of tasks. Each agent possesses multiple possible capacities to correspond to the varying capacity requirements of the different tasks. The competency of agents pertaining to each capability is randomly assigned a value of 0--10. The basic parameter settings for the tasks and agents are presented in Table \ref{T0}.

\begin{table}[!ht]
    \centering
    \setlength{\belowcaptionskip}{0.2cm}
    \caption{Basic parameters of the experimental scenarios.}
    \begin{tabular}{ll}
    \toprule[1.5pt]
        Parameter Description & Value \\ \hline
        Number of feasible tasks per agent & Random in ($0.1|T|$, $0.2|T|$) \\ 
        Number of capacities per agent & Random in [1, 10] \\ 
        Number of capacities per task & Random in [5, 10] \\ 
        Competency value & Random in [1, 10] \\ 
        Cost value of each agent & Random in (1, 20) \\ \bottomrule[1.5pt]
    \end{tabular}
    \label{T0}
\end{table}

Based on these experimental settings, different cost values are assigned when an agent performs different tasks. Three simulation instances under different indicators are designed, including (1) instances under different numbers of agents, (2) instances under different budget rates, and (3) instances under different degrees of heterogeneous cost. Budget rates ${{\alpha }_{s}}$ are introduced to denote the different degrees of budget flexibility for instance $s$, which is denoted as the ratio of the budget to the number of tasks, ${{\alpha }_{s}}={B}/{\left| T \right|}\;$. The degree of cost heterogeneity ${{\gamma }_{s}}$ quantifies the variation in costs associated with different tasks performed by an agent, which is determined by 

\begin{equation}
    {{\gamma }_{s}}=\frac{c_{s}^{\max }-c_{s}^{\min }}{\underset{s}{\mathop{\max }}\,c_{s}^{\max }}
\end{equation}

where $c_{s}^{\max }$ and $c_{s}^{\min }$ represent the maximum and minimum possible cost values and $\underset{s}{\mathop{\max }}\,c_{s}^{\max }$ is the maximum cost value across all scenarios. For instance, considering a specific scenario $s$ in which the highest task cost among all feasible tasks for an agent is 13, the lowest task cost is 7, and the overall maximum cost value across all scenarios is 20, the degree of cost heterogeneity would be $(13-7)/20=0.3$.

\subsection{Ablation experiments}
\noindent
 We conducted ablation experiments using the LLH algorithm, comparing the results of LLH without the CE strategy (abbreviated as LLH-NCE) and LLH without the HLL strategy (abbreviated as LLH-NHL) with the results of the full LLH algorithm. Table \ref{T1} presents the results of the three algorithms in six scenarios with agent scales ranging from 150--900, with each scenario executed independently 10 times. The ``Best'', ``Worst'', and ``Average'' columns represent the best/worst/average results achieved by the algorithm over the 10 runs. ``Gap'' denotes the percentage difference between the algorithm and the full LLH algorithm in terms of the average objective value, ``CU rate'' represents the cost utilization rate of the final coalition solution, where a higher utilization rate indicates a better solution, and ``CPU time'' represents the computation time of the algorithm.

\begin{table*}[!ht]
\centering
\setlength{\belowcaptionskip}{0.2cm}
\caption{Results of LLH, LLH-NHL, and LLH-NCE algorithms executed 10 times with different task scales.}\label{T1}
\begin{tabular}{llllllll}
\toprule[1.5pt]
Agent Num          & Algorithm & Best & Worst & Average & Gap     & CU   rate & CPU   time (s) \\ \hline
\multirow{3}{*}{150} & LLH-NCE   & 806  & 801   & 803.5   & 28.80\% & 97.24\%   & \textbf{0.06}           \\
                     & LLH-NHL   & 998  & 980   & 992.5   & 4.27\%  & 98.48\%   & 0.16           \\
                     & LLH       & \textbf{1044} & \textbf{1014}  & \textbf{1034.9}  & \textbf{0}       & \textbf{98.96\%}   & 0.11           \\ \hline
\multirow{3}{*}{300} & LLH-NCE   & 2380 & 2191  & 2278.2  & 20.73\% & 99.04\%   & \textbf{0.24}           \\
                     & LLH-NHL   & 2564 & 2528  & 2545.5  & 8.05\%  & \textbf{99.58\%}   & 0.63           \\
                     & LLH       & \textbf{2851} & \textbf{2661}  & \textbf{2750.5}  & \textbf{0}       & 99.52\%   & 0.52           \\ \hline
\multirow{3}{*}{450} & LLH-NCE   & 4519 & 3911  & 4040.2  & 13.25\% & 99.33\%   & \textbf{0.64}           \\
                     & LLH-NHL   & 4545 & 4386  & 4430.8  & 3.27\%  & 99.47\%   & 3.06           \\
                     & LLH       & \textbf{4635} & \textbf{4523}  & \textbf{4575.7}  & \textbf{0}       & \textbf{99.59\%}   & 1.25           \\ \hline
\multirow{3}{*}{600} & LLH-NCE   & 5583 & 5475  & 5530.7  & 12.64\% & 99.01\%   & \textbf{1.12}           \\
                     & LLH-NHL   & 5994 & 5926  & 5965.5  & 4.43\%  & 99.24\%   & 2.98           \\
                     & LLH       & \textbf{6284} & \textbf{6194}  & \textbf{6229.8}  & \textbf{0}       & \textbf{99.59\%}   & 2.14           \\ \hline
\multirow{3}{*}{750} & LLH-NCE   & 7148 & 6984  & 7047.8  & 12.17\% & 99.28\%   & \textbf{1.93}           \\
                     & LLH-NHL   & 7564 & 7483  & 7522.6  & 5.09\%  & 99.73\%   & 4.54           \\
                     & LLH       & \textbf{7935} & \textbf{7877}  & \textbf{7905.5}  & \textbf{0}       & \textbf{99.82\%}   & 3.81           \\ \hline
\multirow{3}{*}{900} & LLH-NCE   & 8118 & 7862  & 7998    & 13.88\% & 98.77\%   & \textbf{10.72}          \\
                     & LLH-NHL   & 8813 & 8699  & 8752.7  & 4.06\%  & 99.18\%   & 15.56          \\
                     & LLH       & \textbf{9183} & \textbf{9052}  & \textbf{9107.9}  & \textbf{0}       & \textbf{99.83\%}   & 11.19          \\ \bottomrule[1.5pt]
\end{tabular}
\end{table*}

In terms of the objective values, LLH outperforms both LLH-NCE and LLH-NHL in all scenarios with varying numbers of agents. This illustrates the effectiveness of the CE strategy and the HLL strategy designed for budget-constrained task allocation problems with heterogeneous costs. Although the CE strategy leads to a slight increase in the CPU time (which gradually diminishes as the problem scale increases), its significant impact on improving the solution quality and cost utilization rate is noteworthy. Specifically, the results demonstrate that LLH exhibits an average objective value improvement of over 10\% compared with LLH-NCE for each scenario, along with an approximate 1\% increase in the CU rate.

The role of the CE strategy is to facilitate the mutual exchange of agents from two coalitions, thus maximizing the objective value and the CU rate. In contrast, LLH-NCE only allows agents to perform coalition selection, with no possibility of cooperative exchanges. This limitation results in scenarios where, even if an unallocated agent has lower cost values and higher performance function values, the agent cannot be assigned to tasks because of budget constraints. This highlights the significance of the CE strategy in optimizing the task allocation process. By permitting cooperative exchanges among agents, higher-quality solutions can be achieved.

The HLL strategy does not produce as significant an improvement in solution quality as the CE strategy (approximately 5\%), although it effectively reduces the CPU time by at least 20\%. This can be attributed to the consideration of cost-effectiveness in each iteration of the HLL algorithm, rather than simply choosing the action with the highest utility value. Additionally, the design of the log-linear learning mechanism balances exploitation against exploration during the early and late stages of the solving process, thereby enhancing the overall solving efficiency.

\begin{figure}[!htb]
    \centering
    \subfigure[LLH-NCE]{
        \includegraphics[scale=0.4]{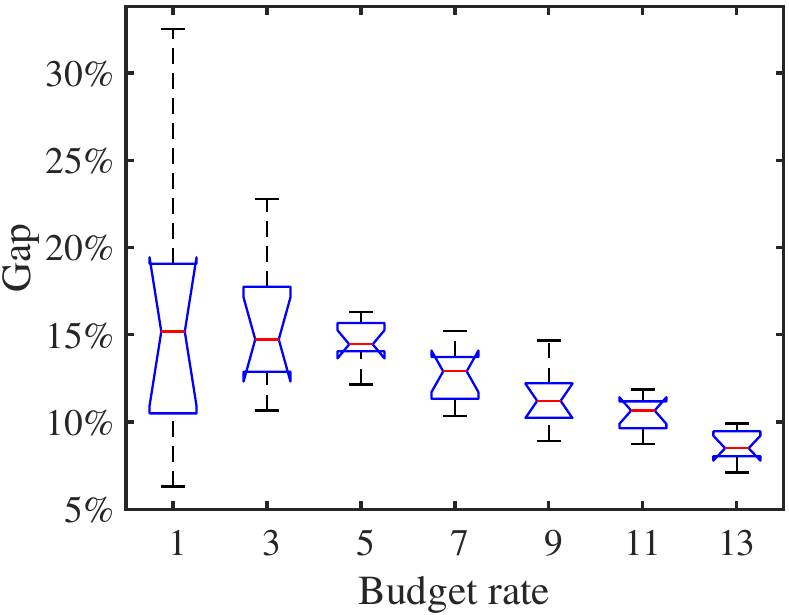}
    \label{F3A}
    }
    \quad
    \subfigure[LLH-NHL]{
        \includegraphics[scale=0.4]{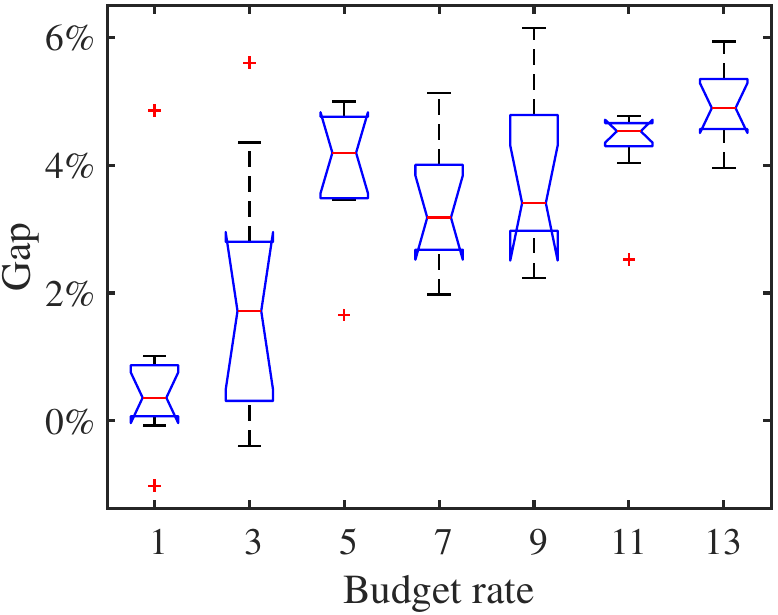}
    \label{F3B}
    }
\caption{LLH gain over LLH-NCE and LLH-NHL with 600 agents at different budget rates ${{\alpha }_{s}}$.}
\label{f3}
\end{figure}

We also explored the enhancements produced by the designed strategies under varying levels of budget flexibility (Budget Rate) and heterogeneous costs (Heterogeneous-cost Degree). The experiments were conducted with 600 agents. Figure 3 illustrates the gap between LLH-NCE and LLH-NHL compared with LLH at different budget rates ${{\alpha }_{s}}$, where the y-axis depicts the objective gain, defined in Table \ref{T1} as the ``Gap''. As shown in Fig. \ref{f3}, the CE strategy performs better at a lower budget rate ${{\alpha }_{s}}$, while HLL assists the algorithm in exploring higher returns when ${{\alpha }_{s}}>5$.

\begin{figure}[!htb]
    \centering
    \subfigure[LLH-NCE]{
        \includegraphics[scale=0.4]{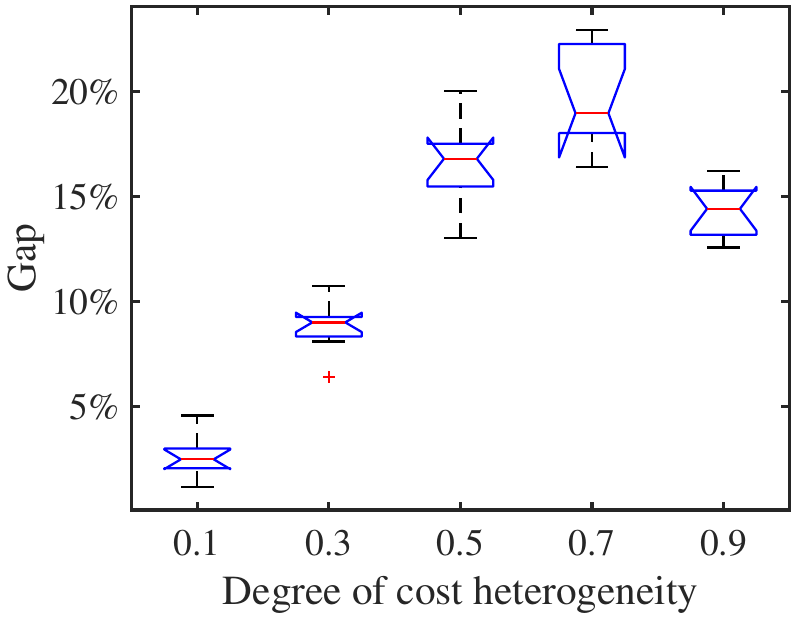}
    \label{F4A}
    }
    \quad
    \subfigure[LLH-NHL]{
        \includegraphics[scale=0.4]{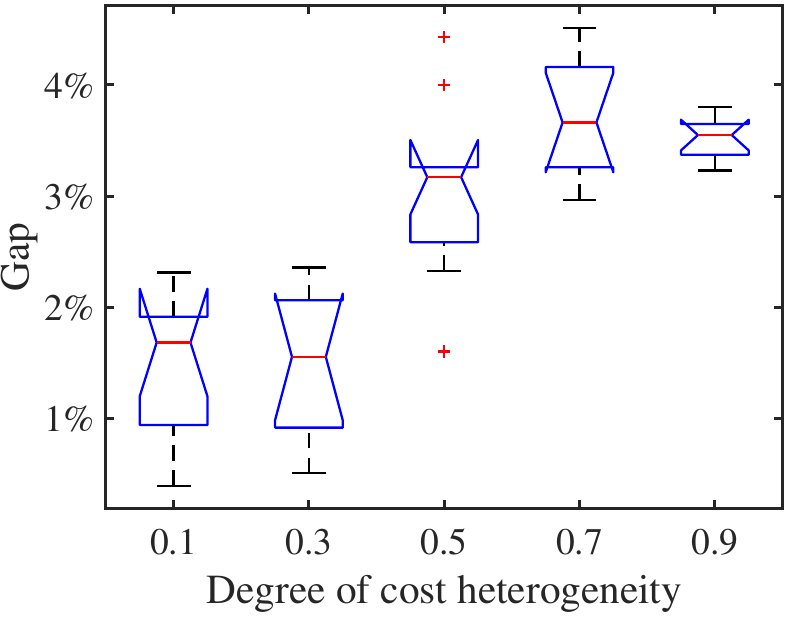}
    \label{F4B}
    }
\caption{LLH gain over LLH-NCE and LLH-NHL with 600 agents at different degrees of cost heterogeneity.}
\label{f4}
\end{figure}

Furthermore, as shown in Fig. \ref{f4}, the difference between LLH-NCE and LLH-NHL compared with LLH gradually widens as the heterogeneous-cost degree increases. This demonstrates the effectiveness of the CE strategy and HLL strategy in addressing greater heterogeneity of costs in the allocation system.

\subsection{Algorithm comparison experiments}
\noindent
This subsection evaluates the effectiveness of the proposed LLH under various agent scales, budget rates, and degrees of cost heterogeneity. The results are compared with those given by several distributed task allocation algorithms and potential game theory algorithms.

$\bullet$ Cost-efficiency factor algorithm (CF)\cite{7}: The CF algorithm uses a cost-efficiency factor to guide the coalition joined by an unassigned agent. At each iteration, the agent and the action pair with the global maximum value are selected until there is no feasible action.\footnote{To make this algorithm applicable to our problem of a heterogeneous cost for each task, we replace the cost of the agent in the original cost-efficiency factor with the average cost of the agent performing different tasks.}

$\bullet$ Better reply process (BRP)\cite{37}: In BRP, each agent maintains a better reply action set ${{B}_{i}}(a)=\{{{{a}'}_{i}}\in {{A}_{i}}\left| {{U}_{i}}({{{{a}'}}_{i}},{{a}_{-i}})>{{U}_{i}}(a) \right.\}$, which contains all actions that obtain a greater utility value than the current action. For any time $t>0$, agents continue with the current action with probability  $\chi$ or update the action selected from ${{B}_{i}}(a)$ with probability $1-\chi $.

$\bullet$ Best response algorithm (BRA)\cite{36}: Different from BRP, this is a multi-round greedy algorithm. At each iteration, one agent is randomly chosen and allowed to select the best action from the constrained action set.

The comparison results using different numbers of agents are presented in Table \ref{T2}.

\begin{table*}[!htb]
\centering
\setlength{\belowcaptionskip}{0.2cm}
\caption{Comparison results over 10 repeated executions with different task scales.}\label{T2}
\begin{tabular}{llllllll}
\toprule[1.5pt]
Agent   Num          & Algorithm & Best & Worst & Average & Gap     & CU   rate & CPU   time (s) \\ \hline
\multirow{4}{*}{50}  & CF        & 996  & 979   & 986.9   & 4.86\%  & \textbf{99.60\%}   & 0.12           \\
                     & BRP       & 1113 & 908   & 1010.5  & 2.41\%  & 98.96\%   & 0.15           \\
                     & BRA       & 1016 & 998   & 1006.8  & 2.79\%  & 97.32\%   & 0.13           \\
                     & LLH       & \textbf{1044} & \textbf{1014}  & \textbf{1034.9}  & \textbf{0.00\%}  & 98.56\%   & \textbf{0.11}           \\ \hline
\multirow{4}{*}{100} & CF        & 2495 & 2457  & 2477    & 11.04\% & 98.20\%   & 1.65           \\
                     & BRP       & 2796 & 2552  & 2721.4  & 1.07\%  & \textbf{99.56\%}   & 0.74           \\
                     & BRA       & 2623 & 2567  & 2597    & 5.91\%  & 99.24\%   & \textbf{0.50}           \\
                     & LLH       & \textbf{2851} & \textbf{2661}  & \textbf{2750.5}  & \textbf{0.00\%}  & 99.52\%   & 0.52           \\ \hline
\multirow{4}{*}{150} & CF        & 4010 & 3955  & 3981.6  & 14.92\% & 98.93\%   & 7.42           \\
                     & BRP       & 4612 & 4359  & 4509.6  & 1.47\%  & 99.49\%   & 1.90           \\
                     & BRA       & 4529 & 4446  & 4487.2  & 1.97\%  & 98.56\%   & \textbf{1.16}           \\
                     & LLH       & \textbf{4635} & \textbf{4523}  & \textbf{4575.7}  & \textbf{0.00\%}  & \textbf{99.97\%}   & 1.25           \\ \hline
\multirow{4}{*}{200} & CF        & 5487 & 5419  & 5452.6  & 14.25\% & 99.40\%   & 22.60          \\
                     & BRP       & 6184 & 6001  & 6095.9  & 2.20\%  & 99.54\%   & 3.40           \\
                     & BRA       & 6116 & 6032  & 6065.8  & 2.70\%  & 99.21\%   & \textbf{1.82}           \\
                     & LLH       & \textbf{6284} & \textbf{6194}  & \textbf{6229.8}  & \textbf{0.00\%}  & \textbf{99.92\%}   & 2.14           \\ \hline
\multirow{4}{*}{250} & CF        & 7145 & 7077  & 7114.2  & 11.12\% & 99.28\%   & 53.87          \\
                     & BRP       & 7874 & 7676  & 7758.4  & 1.90\%  & 99.54\%   & 7.41           \\
                     & BRA       & 7759 & 7620  & 7678.5  & 2.96\%  & 99.78\%   & \textbf{3.72}           \\
                     & LLH       & \textbf{7935} & \textbf{7877}  & \textbf{7905.5}  & \textbf{0.00\%}  & \textbf{99.80\%}   & 3.81           \\ \hline
\multirow{4}{*}{300} & CF        & 8609 & 8502  & 8567.4  & 6.31\%  & 99.47\%   & 214.99         \\
                     & BRP       & 9059 & 8856  & 8937.4  & 1.91\%  & 99.62\%   & 17.15          \\
                     & BRA       & 8839 & 8605  & 8750.1  & 4.09\%  & 99.33\%   & \textbf{10.51}          \\
                     & LLH       & \textbf{9183} & \textbf{9052}  & \textbf{9107.9}  & \textbf{0.00\%}  & \textbf{99.83\%}   & 12.19          \\ \bottomrule[1.5pt]
\end{tabular}
\end{table*}

The proposed LLH clearly performs best in terms of the objective values with all numbers of agents, and obtains the best cost rate in most instances. The modified heterogeneous-cost CF algorithm performs relatively poorly in terms of the CPU time and solution quality, possibly because the original CF algorithm is not designed to solve heterogeneous cost problems. By directly averaging the heterogeneous cost of each agent, it is difficult for CF to accurately estimate the cost-efficiency factor of each action, which causes unfeasible actions to be selected.

Among the three potential game theory algorithms (i.e., BRA, BRP, and LLH), although BRA takes less time to calculate a solution through the introduction of pure greediness, it gives the worst solution quality owing to a lack of randomness. In contrast, because of the randomness in the action selection from the better reply set, BRP obtains a slightly better average solution than BRA, but requires a longer computation time. The proposed LLH algorithm lies somewhere in between, maintaining a balance of greediness and randomness that performs well in early iterations and obtains the best final performance.

Based on analysis of the algorithms under different budget rates and heterogeneous cost degrees, Fig. 5 compares the objective function values obtained by the algorithms for budget rates ranging from 1--13. The y-axis represents the improvement in the objective value compared with the worst-case objective value under the same budget rate. It is evident that, as the budget rate increases, the advantage of LLH over the other algorithms becomes more pronounced. This can be attributed to the expansion of the action space for each agent as the budget rate increases. The log-linear learning mechanism is more advantageous in solving the action selection problem in the context of a larger action space. This is because it can effectively control the learning parameters to explore the possibility of a better solution in the early iterations, before subsequently exploiting and improving the existing solution. Thus, the relative performance of LLH improves as the budget rate increases.

\begin{figure}[!htb]
    \centering
    \includegraphics[scale=0.45]{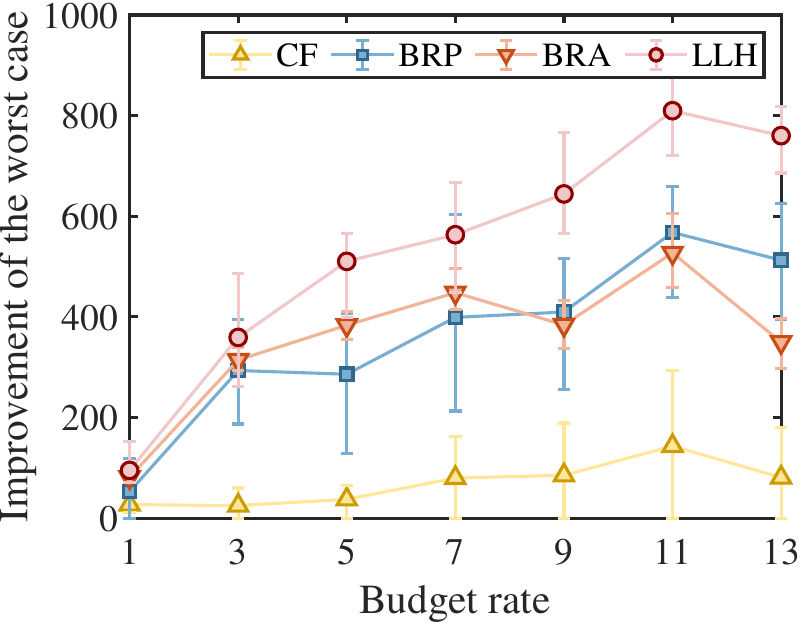}  
\caption{Algorithm comparison results at various budget rates.}   
\label{f5}
\end{figure}

Figure 6 compares the algorithms under different heterogeneous cost degrees. The x-axis represents the heterogeneous cost degree, indicating a greater disparity in the costs of executing different tasks for the same agent. The proposed LLH algorithm exhibits a significant objective improvement as the heterogeneous cost degree increases. This highlights the advantages and adaptability of LLH in addressing task allocation problems with budget constraints under varying heterogeneous costs. As the heterogeneous cost degree decreases, the performance of the CF algorithm becomes slightly better than that of the proposed LLH algorithm. This can be attributed to the ability of CF to provide cost estimates close to the actual values for each task when the heterogeneity is low. 

\begin{figure}[!htb]
    \centering
    \includegraphics[scale=0.45]{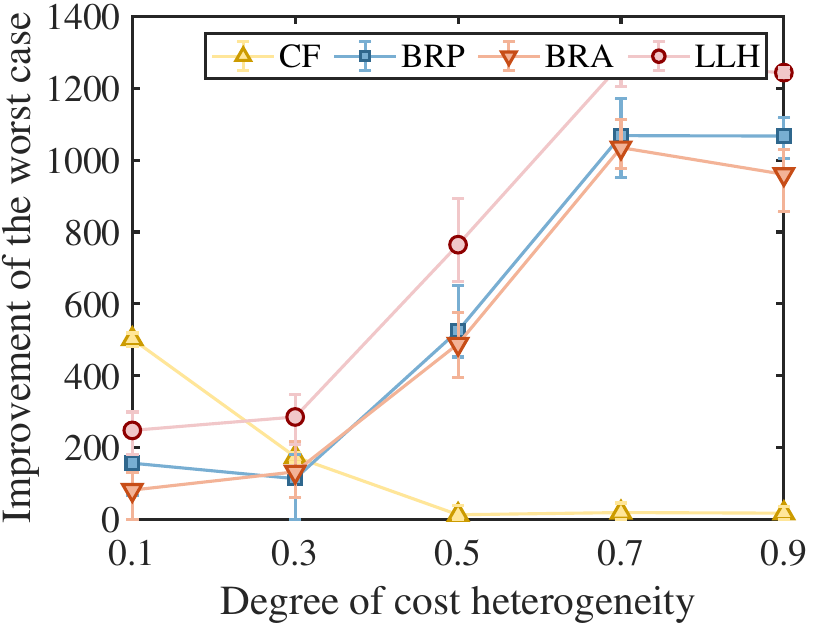}  
\caption{ Algorithm comparison result with various degrees of cost heterogeneity.}   
\label{f6}
\end{figure}

\section{Conclusions}
This paper has addressed the HCTAB problem, in which a set of heterogeneous agents is allocated to a set of tasks under a budget constraint. Our approach is a pioneering attempt to address the HCTAB problem from a game-theoretic perspective, exploring a more efficient, distributed, and convergence-guaranteed solution framework. The proposed LLH algorithm exhibits a significant objective improvement as the heterogeneous cost degree increases. This highlights the advantages and adaptability of LLH in addressing task allocation problems with budget constraints under varying levels of heterogeneous cost. Extensive empirical studies have illustrated the effectiveness and superiority of the LLH algorithm in terms of the solution quality across various agent scales, budget rates, and (most) degrees of cost heterogeneity.

Future studies will consider the HCTAB problem in dynamic scenarios, exploring how to assign tasks to agents when the rewards for performing heterogeneous tasks change dynamically. To adapt to this dynamic paradigm, a deep Q-learning method could be added to the existing LLH algorithm to predict the reward changes and achieve a data-driven distributed task allocation method.

\vskip 2mm
\zihao{5}
\noindent
\textbf{Acknowledgment}
\vskip 2mm

\zihao{5--}
\noindent
This research was supported by the National Natural Science Foundation of China (72001212, 72201272) and the Young Elite Scientists Sponsorship Program by CAST (2022QNRC001).

\vskip 2mm

\renewcommand\refname{\zihao{5}\textbf{References}}

\bibliographystyle{ieeetr}

\bibliography{ref_new.bib}




\end{document}